\documentclass[11pt]{article}

\begin{document}
	\def\ba{\begin{eqnarray}}
	\def\ea{\end{eqnarray}}
	\def\w{\wedge}
	\def\d{\mbox{d}}
	\def\D{\mbox{D}}

\begin{titlepage}	
\title{Non-Riemannian Description of  Robinson-Trautman Spacetimes  in Brans-Dicke Theory Of Gravity}
\author{ Muzaffer Adak\footnote{madak@pau.edu.tr} 
\\ {\small Department of Physics, Pamukkale University, 20017 Denizli, Turkey} \\
\\ Tekin Dereli\footnote{tdereli@ku.edu.tr}    
\\
{\small  Department of Physics, Ko\c{c} University, 34450 Sar{\i}yer-\.{I}stanbul, Turkey} \\ 
\\
Yorgo \c{S}eniko\u{g}lu\footnote{ysenikoglu@ku.edu.tr}
\\
{\small  Department of Physics, Ko\c{c} University, 34450 Sar{\i}yer-\.{I}stanbul, Turkey}  }

\vskip 1cm 

\date{ }

\maketitle	
	
\vskip 2cm 

\begin{abstract}
\noindent The variational field equations of Brans-Dicke scalar-tensor theory of gravitation are given in a non-Riemannian setting in the language of exterior differential forms 
over  4-dimensional spacetimes. A conformally re-scaled Robinson-Trautman metric together with the Brans-Dicke scalar field are used to characterise
algebraically special Robinson-Trautman spacetimes. All the relevant tensors are worked out in a complex null basis  and given explicitly in an appendix
for future reference. Some special families of solutions are also given and discussed.

\vskip 2cm

\end{abstract}
\end{titlepage}

\newpage

 \section{Introduction}

Robinson-Trautman spacetimes \cite{robinson-trautman1,robinson-trautman2} were first presented in the early 60's. Since then, they became one of the most important solutions of general relativity in our understanding of gravitational waves. These physically important solutions to the Einstein field equations are found by considering the fact that they should admit a null, geodesic and congruence without shear or twist. In order to illustrate the physical meaning of these solutions of the empty space Einstein field equations, it is found useful to study the asymptotic invariants of the Weyl tensor\cite{exact solutions}.  Mathematical studies on the global structure of Robinson-Trautman spacetimes are presented vividly in the literature\cite{chrusciel}, including cases of radiative space-times with cosmological constant\cite{bicak-podolsky,podolsky-svitek2} using analytical methods. One may consult the relevant chapters in Ref.\cite{griffiths-podolsky} for details.
 More recent works\cite{podolsky-svarc} on the algebraic classification of Robinson-Trautman spacetimes provide non-trivial explicit examples in which the focus is on  specific algebraic properties of the Robinson-Trautman spacetimes with a free scalar field.  We note that the Robinson-Trautman spacetimes are frequently used to describe radially propagating spherical gravitational waves. Such studies are specifically realistic to depict imploding spherical stars and/or coalescing binary black holes.

 Brans-Dicke scalar-tensor theory of gravity also dates from the early1960's and it has been recognised very early as the most viable alternative to Einstein's theory of general relativity. It has found many applications since then  to problems in gravitation and cosmology. String theorists have also pointed out similarities between string models and scalar-tensor theories, prompting further investigation of the latter\cite{guven}. Actually, the gravitational coupling constant has a characteristic aspect in modern cosmology: It may depend on the cosmic time. This idea is initially due to Dirac\cite{dirac} who noticed that the analysis of constants of cosmological nature and the fundamental physical constants will become inter-related if these so-called constants are allowed to vary slowly  over cosmological time scales. The work of Brans and Dicke \cite{brans-dicke,dicke,brans1} accomplished to detail this idea where a scalar field is regarded as a local gravitational coupling and carries gravity together with the metric tensor of general relativity. On the one hand, G\"{u}rsey\cite{gursey} argued that Einstein's theory of general relativity as it stands incorporates Mach's Principle and detailed the reformulation of the metric as a product of a scalar density $\phi^2$ and a tensor density. On the other hand, conformal re-scaling properties of scalar-tensor gravity theories have been further elaborated, for instance, by Deser\cite{deser} and Anderson\cite{anderson}.Here, we are generalising this model by considering a non-Riemannian description of spacetime where the spacetime torsion is determined by the 
gradient of a  scalar field\cite{dereli1,obukhov}. We are going to  include in our discussion an arbitrary  conformal scale factor of the metric as well 
in order to analyse the solutions in different "frames" (pictures).  
Robinson-Trautman solutions in the Einstein "frame" (picture) have been studied much in detail while the corresponding solutions in Brans-Dicke-Jordan "frame" (picture) or in the string "frame" (picture) are 
physically equivalent to those in the Einstein "frame" (picture) in the sense that these pictures could be related  to each other  by certain re-scalings of the basic fields. 
However, since a non-Riemannian description of scalar-tensor theories in the latter two pictures are possible\cite{dereli1}, the equations of motion of non-spinning test masses in such spacetimes may differ in general from the geodesic equations of motion in a Riemannian setting\cite{dereli2,dereli3,cebeci-dereli-tucker,tucker2}. 
 Motivated by these considerations, we  discuss in this paper the Robinson-Trautman solutions of the Brans-Dicke field equations in a non-Riemannian setting. 
We are going to use extensively the language of exterior differential forms in  a complex null basis.

 The paper is organised as follows. In Section:2, the Brans-Dicke field equations will be derived by a first order variational principle using the language of exterior differential forms. The role played by the space-time torsion 2-forms that are proportional to the gradient of the Brans-Dicke scalar field will be discussed. 
The non-Riemannian space-time geometry expressed in terms of complex null differential forms and related formulas are given separately in the Appendix A.
In Section:3, a  conformally re-scaled  Robinson-Trautman metric ansatz and a corresponding  scalar field assumption are given.  
All the relevant geometrical expressions are explicitly worked out in Appendix B. 
We reduce the Brans-Dicke field equations in Robinson-Trautman spacetimes to a system of coupled non-linear differential equations and check
their vacuum Einstein solutions for consistency. Furthermore a family of static, spherically symmetric solutions 
with non-vanishing scalar charge (the Janis-Newman-Winicour solutions) of the coupled Einstein-massless scalar field equations is written down.
Moreover a new family of propagating solutions is constructed. It corresponds to a special case of recently found Robinson-Trautman type solutions of the Einstein 
massless scalar field equations\cite{tahamtan-svitek1,tahamtan-svitek2}. Section:4 is devoted to concluding remarks. 


\section{Brans-Dicke Gravity}

\noindent  There are strong reasons motivated by recent astrophysical and cosmological observations that Einstein's general relativity theory may require 
the inclusion of certain yet undetected scalar fields of either gravitational or matter origin. Furthermore low energy string field theories involve many unobserved scalar degrees of freedom and the most unified theories of strong and electroweak interactions predict scalar fields with both astrophysical and cosmological implications. In fact, Brans and Dicke suggested in 1961 a modification of Einsteinian gravity by introducing a real scalar field with particular couplings to matter via the space-time metric. This is arguably the simplest 
modification of general relativity. Brans-Dicke theory  is a scalar-tensor theory of gravity  that is still regarded as the most viable alternative to Einstein's theory.  A real parameter
$\omega$ characterises the gravitational effects of the scalar field $\phi$ so that predictions of the Brans-Dicke theory are indistinguishable from those of Einstein's theory for large values of $\omega$. Newton's universal gravitational coupling constant $G$ is usually related in this theory to solutions which have a constant value of the scalar field ,
 i.e. $<\phi> \sim \frac{1}{8\pi G}$. This modified theory can also be invoked in the context of conformal re-scalings of the space-time metric. A  locally scale invariant theory of gravity occurs for the specific case for $\omega=-\frac{3}{2}$. The field equations of the Brans-Dicke theory are derived by a variational principle from an action $I=\int_{M} \mathcal{L},$ where the  action density 4-form
 \begin{equation}
 {\mathcal{L}} = \frac{\phi}{2}R_{ab} \wedge *(e^a \wedge e^b) -\frac{\omega}{2\phi} d\phi \wedge *d\phi + \Lambda \phi^2 *1.
  \end{equation}
 $\{e^a\}$ are the co-frame 1-forms in terms of which the space-time metric is given by $g=\eta_{ab}e^{a} \otimes e^b$ with $\eta_{ab}=diag(-+++)$. ${}^*$ denotes 
the Hodge map referring to the space-time orientation $*1 = e^0 \wedge e^1 \wedge e^2 \wedge e^3$. $\phi$ is the Brans-Dicke scalar field. We also introduced a cosmological constant $\Lambda$. 
$\{{\omega}^{a}_{\;\;b}\}$ are the connection 1-forms that satisfy the Cartan structure equations
  \begin{equation}
  de^a + \omega^{a}_{\;\;b}  \wedge e^b = T^a
  \end{equation}
  with the torsion 2-forms $T^a$ and  
  \begin{equation}
 d \omega^{a}_{\;\;b} + \omega^{a}_{\;\;c}  \wedge \omega^{c}_{\;\;b} =R^{a}_{\;\;b}   
  \end{equation}
are the curvature 2-forms of spacetime. 
Conventionally the Levi-Civita connection 1-forms 
$\{{\hat{\omega}}^{a}_{\;\;b}\}$ are assumed which are fixed in a unique way by the metric tensor through the Cartan structure equations
\begin{equation}
de^a + {\hat{\omega}}^{a}_{\;\;b} \wedge e^b = 0.
\end{equation}
Now the connection 1-forms admit a unique decomposition,
 \begin{equation}
 \omega^{a}_{\;\;b} = {\hat{\omega}}^{a}_{\;\;b} + K^{a}_{\;\;b} 
 \end{equation}
 in terms of the contortion 1-forms $K^{a}_{\;\;b} $ that satisfy  $K^{a}_{\;\;b} \wedge e^b = T^a$. 
Then the field equations derived from the action above varied independently with respect to the co-frames and the scalar field, 
subject to the constraint that the  connection is Levi-Civita, turn out to be $(\omega \neq -\frac{3}{2})$
\begin{eqnarray}
-\frac{\phi}{2}{\hat{R}}^{bc} \wedge *(e_a \wedge e_b \wedge e_c) &=& \frac{\omega}{2\phi} \left (  \iota_a d\phi *d\phi + d\phi \wedge \iota_a *d\phi  \right ) \nonumber \\
& & +\hat{D}(\iota_a*d\phi) +\Lambda \phi^2 *e_a, \nonumber \\
d*d\phi =0 & &
\end{eqnarray}
where $\hat{D}$ denotes the covariant exterior derivative with respect to the Levi-Civita connection. 
A first order variational derivation of these field equations is also possible with some interesting consequences. In that case, we set $\phi = \alpha^2$ for convenience. 
Then the Brans-Dicke action density 4-form becomes
\begin{equation}
 {\mathcal{L}} = \frac{\alpha^2}{2}R_{ab} \wedge *(e^a \wedge e^b) -\frac{c}{2} d\alpha \wedge *d\alpha + \Lambda \alpha^4 *1,
  \end{equation}
  where $c$ is a real coupling constant. This action will be varied independently with respect to the co-frame 1-forms $e^a$, connection 1-forms $\omega^{a}_{\;\;b}$ and
  the scalar field $\alpha$. The corresponding coupled field equations turn out to be $(c \neq 0)$:
  \begin{eqnarray}
 -\frac{\alpha^2}{2} R^{bc} \wedge *(e_a \wedge e_b \wedge e_c) &=& c\hspace{1mm}\tau_a[\alpha] +\Lambda \alpha^4 *e_a, \nonumber \\ 
  T^a &=& e^a \wedge \frac{d \alpha}{\alpha} ,   \nonumber \\ 
    c d*d\alpha^2 &=& 0; 
 \end{eqnarray}
  where
  \ba
  \tau_a[\alpha]=\frac{1}{2} \left (  \iota_a d\alpha *d\alpha + d\alpha\wedge \iota_a *d\alpha  \right ) \equiv T_{ab} *e^b
  \ea
  are the energy-momentum 3-forms of the scalar field.
 Therefore, taking the torsion 2-forms given by the above variational field equations and after some algebra, 
 we arrive at the following decomposition for the Einstein 3-forms
\begin{eqnarray}
R^{bc} \wedge *(e_a \wedge e_b \wedge e_c) = & & {\hat{R}}^{bc} \wedge *(e_a \wedge e_b \wedge e_c) -\frac{4}{\alpha^2} \left (  \iota_a d\alpha *d\alpha + 
d\alpha \wedge \iota_a *d\alpha  \right ) \nonumber \\ & & - \frac{2}{\alpha^2} \iota_a (d\alpha \wedge *d\alpha ) + \frac{4}{\alpha} {\hat{D}} (*d\alpha) .
\end{eqnarray} 
 Substituting these  expressions in the variational field equations we find
\begin{eqnarray}
-\frac{\alpha^2}{2}{\hat{R}}^{bc} \wedge *(e_a \wedge e_b \wedge e_c) &=& \frac{(c-6)}{2} \left (  \iota_a d\alpha *d\alpha + d\alpha \wedge \iota_a *d\alpha  \right ) \nonumber \\
& & +\hat{D}(\iota_a*d\alpha^2) +\Lambda \alpha^4 *e_a,  \nonumber \\
c d*d\alpha^2 &=&0.
\end{eqnarray} 
  These coincide with the Brans-Dicke equations given before for $\phi=\alpha^2$ provided we make the identification 
  \begin{equation}
  c=4\omega+6.
  \end{equation} 
We also note the conformal re-scalings of the metric induced by the local transformations
\begin{equation}
g \rightarrow e^{2\sigma(x)} g.
\end{equation}
Then 
\begin{equation}
e^a \rightarrow e^{\sigma} e^a , \quad    {\hat{\omega}}^{a}_{\;\;b} \rightarrow {\hat{\omega}}^{a}_{\;\;b} -(\iota^a d\sigma) e_b +  (\iota_b d\sigma) e^a .
\end{equation}
If we now postulate the conformal re-scaling of the scalar field as
\begin{equation}
\alpha \rightarrow e^{-\sigma} \alpha,
\end{equation}
then it follows that
\begin{equation}
K^{a}_{\;\;b} \rightarrow K^{a}_{\;\;b} +(\iota^a d\sigma) e_b -  (\iota_b d\sigma) e^a .
\end{equation}
Therefore we have  $\omega^{a}_{\;\;b}  \rightarrow \omega^{a}_{\;\;b}$ and $R^{a}_{\;\;b}  \rightarrow R^{a}_{\;\;b}$, so that the Brans-Dicke action density 4-form 
is conformally scale invariant for the choice $c=0$, that is, for $\omega =-\frac{3}{2}$. 
In fact, Brans and Dicke postulated  independently of their field equations that the non-spinning test masses should follow geodesic equations of motion given by
 \ba
{\hat{\nabla}}_{\dot{C}}\dot{C}= 0,  
\ea
where $\hat{\nabla}$ is the unique Levi-Civita connection of $g$ and $\dot{C}$ is a unit,time-like tangent vector field so that $g(\dot{C},\dot{C})=-1.$
On the other hand  here
the autoparallel equations of motion relative to the non-Riemannian connection $\nabla$ can be decomposed as
\ba
\nabla_{\dot{C}}\dot{C}= {{\hat{\nabla}}}_{\dot{C}} \dot{C} + \frac{1}{\alpha} \iota_{\dot{C}}(d\alpha \wedge \widetilde{\dot{C}} ) =0,  
\ea
and would differ in general from the geodesic equations of motion.

\noindent Once the conformal re-scaling properties of the connection and curvature are set, 
a re-formulation of the Brans-Dicke action is possible if we adopt another orthonormal frame defined by ${\tilde{e}}_a = \frac{\alpha}{\alpha_0} e^a$ where $\alpha_0$ is a constant
that fixes a scale. Note that since $e^a$'s are orthonormal with respect to the metric $g$, we must require ${\tilde{e}}^a$'s to be orthonormal with respect to a re-scaled metric
\ba 
\tilde{g} = \left ( \frac{\alpha}{\alpha_0}  \right )^2 g.
\ea 
Then it can be shown that our action density 4-form takes the form
\ba
 {\mathcal{L}} = \frac{{\alpha_0}^2}{2} R_{ab} \wedge \tilde{*}(\tilde{e}^a \wedge \tilde{e}^b) -\frac{c {\alpha_0}^2}{2 \alpha^2} d\alpha \wedge \tilde{*}d\alpha +
 \alpha_{0}^{4}\Lambda \tilde{*}1.
\ea 
For $c \neq 0$, a further re-definition
\ba
\Phi = \ln \left ( \frac{\alpha}{\alpha_0}  \right ),
\ea
gives the action density 4-form in the Einstein picture:
\ba
 {\mathcal{L}} = \frac{{\alpha_0}^2}{2} R_{ab} \wedge \tilde{*}(\tilde{e}^a \wedge \tilde{e}^b) -\frac{c {\alpha_0}^2}{2} d\Phi \wedge \tilde{*}d\Phi +
 \alpha_{0}^{4}\Lambda \tilde{*}1.
\ea 
The connection variations of this action yield the equation
\ba
D_{\omega}\tilde{*}({\tilde{e}}^a \wedge {\tilde{e}}^b ) =0
\ea
whose unique solution implies that the connection should be the  torsion-free Levi-Civita connection of the metric $\tilde{g}$.  Thus although the torsion associated with the fields 
($g, \alpha$) in Brans-Dicke-Jordan picture need not be zero, the torsion of the re-defined fields  in  Einstein picture ($\tilde{g}, \Phi$) should be zero. 
In particular, we wish to emphasise that the auto-parallel curves determined in terms of the connection with torsion in the Brans-Dicke-Jordan picture are equivalent 
to geodesic curves in the Einstein picture. This is essentially the observation of Dirac. On the other hand, Brans and Dicke argued in their paper that the scalar field $\phi$ should not couple to matter fields. This assumption in fact amounts to postulating geodesic equations of motion for spin-less test particles in the Brans-Dicke-Jordan picture
that would be different in general from the auto-parallel  equations of motion. 



\section{Robinson-Trautman Spacetimes}

\noindent In order to solve the Brans-Dicke field equations we generalise (for convenience) the Robinson-Trautman ansatz as follows\footnote{The separation of variables in the angular part is  different than the standard one, e.g. that one may found in Ref.[3], motivated by some recent classical solutions in this class.}:
\begin{eqnarray}
g &=& e^{2\sigma(u,r)}\hspace{1mm}\left \{ -C(u,r,\zeta,\bar{\zeta}) \d u^2 + 2\d u \d r + \frac{4h^2(u,r)}{P^2(\zeta ,
	\bar{\zeta})} \d \zeta \d \bar{\zeta}\right \}, \nonumber \\
	\alpha&=& \alpha(u,r),   \label{eq:R-Tmetrik0}
\end{eqnarray}
in a coordinate chart $(u,r,\zeta,\bar{\zeta})$ where $u$ is a null coordinate and $\zeta,\bar{\zeta}$ are the stereographic projection coordinates on $S^2$. 
We introduced a new function $h(u,r)$ besides $C(u,r,\zeta,\bar{\zeta})$  and an arbitrary scale factor $\sigma(u,r)$, that all to be determined by the field equations.



\medskip

\noindent Inserting the expressions  for curvatures given in Appendix B  in (70) and using (72) for the stress-energy-momentum tensor,
the Einstein field equations to be solved reduce to the following system:
\ba
&& \Bigg(\frac{h_rC_r}{h}+\frac{h_{rr}C}{h}+\frac{4\sigma_rh_u}{h}+\frac{4\sigma_r\alpha_u}{\alpha}+\frac{\alpha_{rr}C}{\alpha}+\frac{4\sigma_u h_r}{h}+\frac{C_r\alpha_r}{\alpha} \nonumber \\
&&+\frac{4\alpha_r\sigma_u}{\alpha}+\frac{\alpha^2_rC}{\alpha^2}+\frac{2\alpha_r\alpha_u}{\alpha^2}+\frac{h^2_rC}{h^2}+\frac{4\alpha_rh_u}{h\alpha}+\frac{4\alpha_uh_r}{h\alpha}+\frac{4\sigma_rh_rC}{h}\nonumber \\
&&+\frac{4\sigma_rC\alpha_r}{\alpha}-\frac{P_{\zeta\bar{\zeta}}P}{h^2}+\frac{P_{\zeta}P_{\bar{\zeta}}}{h^2}+4\sigma_r\sigma_u+2\sigma^2_rC+\frac{2h_{ur}}{h}+\sigma_rC_r\nonumber \\
&&+\frac{2\alpha_{ur}}{\alpha}+2\sigma_{ur}+\frac{4\alpha_rh_rC}{h\alpha}+\sigma_{rr}C+\frac{2h_rh_u}{h^2}\Bigg)=0\; ,\\
&&\Bigg(\sigma^2_r-\frac{\alpha_{rr}}{\alpha}-\sigma_{rr}-\frac{h_{rr}}{h}+\frac{2\sigma_r\alpha_r}{\alpha}+\frac{2\alpha^2_r}{\alpha^2}-\frac{1}{2}\frac{c\alpha^2_r}{\alpha^2}\Bigg)=0\; , \\
&&\Bigg(\frac{C_{r\zeta}}{2}+\frac{C_{\zeta}\alpha_r}{\alpha}+\sigma_rC_{\zeta}\Bigg)=0\; ,\\
&&\Bigg(\frac{P^2C_{\zeta\bar{\zeta}}}{h^2}-\frac{4C\alpha_{ur}}{\alpha}-\frac{2C_u\alpha_r}{\alpha}+\frac{4\sigma_uC\alpha_r}{\alpha}+\frac{4\sigma_rC\alpha_u}{\alpha}+\frac{2\sigma_rC^2\alpha_r}{\alpha}\nonumber \\
&&-\frac{h_{rr}C^2}{h}+\frac{2\alpha^2_rC^2}{\alpha^2}-\frac{\alpha_{rr}C^2}{\alpha}+\frac{2h_uC_r}{h}-4\sigma_{uu}+\sigma^2_rC^2+2\sigma_uC_r\nonumber \\
&&-\sigma_{rr}C^2+\frac{8\alpha_uC\alpha_r}{\alpha^2}+\frac{8\alpha^2_u}{\alpha^2}-\frac{4\alpha_{uu}}{\alpha}+4\sigma^2_u+4\sigma_u\sigma_rC+\frac{8\sigma_u\alpha_u}{\alpha}\nonumber \\
&&+\frac{2\alpha_uC_r}{\alpha}-\frac{4h_{uu}}{h}-4\sigma_{ur}C-2\sigma_rC_u-\frac{4h_{ur}C}{h}-\frac{2h_rC_u}{h}\nonumber\\
&&-\frac{1}{2}\frac{c\alpha^2_rC^2}{\alpha^2}-\frac{2c\alpha^2_u}{\alpha^2}-\frac{2c\alpha_uC\alpha_r}{\alpha^2}\Bigg)=0\; , \\
&&\Bigg(\frac{h_rC_r}{h}+\frac{h_{rr}C}{h}+\frac{2\sigma_rh_u}{h}-\frac{2\sigma_r\alpha_u}{\alpha}+\frac{2\alpha_{rr}C}{\alpha}+\frac{2\sigma_uh_r}{h}+\frac{2C_r\alpha_r}{\alpha}\nonumber \\
&&+\frac{2\alpha_r\sigma_u}{\alpha}+\frac{1}{2}C_{rr}+\frac{2\alpha_rh_rC}{h\alpha}+\frac{2\alpha_rh_u}{h\alpha}+\frac{2\alpha_uh_r}{h\alpha}+\frac{2\sigma_rh_rC}{h}\nonumber \\
&&+\frac{2\sigma_rC\alpha_r}{\alpha}+\frac{\alpha^2_rC}{\alpha^2}+\frac{2\alpha_r\alpha_u}{\alpha^2}+2\sigma_{rr}C+2\sigma_r\sigma_u+\sigma^2_rC+\frac{2h_{ur}}{h}\nonumber \\
&&+2\sigma_rC_r+\frac{4\alpha_{ur}}{\alpha}+4\sigma_{ur}+\frac{c\alpha_u\alpha_r}{\alpha^2}+\frac{1}{2}\frac{c\alpha^2_rC}{\alpha^2}\Bigg)=0\; . 
\ea
We also work out the scalar field equation to complete this system: 
\ba
&& c \Bigg\{ 4\frac{\alpha_u}{\alpha} \frac{h_r}{h}+4 \frac{\alpha_r}{\alpha} \frac{h_u}{h}+4 \frac{\alpha_r}{\alpha} \frac{h_r}{h}C +  2(\frac{\alpha_r}{\alpha})^2 C
+2\frac{\alpha_{rr}}{\alpha}C  \\
&&+2\frac{\alpha_r}{\alpha} C_r+4 \frac{\alpha_r}{\alpha} \frac{\alpha_u}{\alpha} +4\frac{\alpha_{ur}}{\alpha} +4\sigma_r\frac{\alpha_u}{\alpha} +4\sigma_r\frac{\alpha_r}{\alpha} C 
+4 \sigma_u \frac{\alpha_r}{\alpha}\Bigg \} =0.  \nonumber
\ea
Thus we have  five functions $C(u,r,\zeta,\bar{\zeta})$, $P(\zeta,\bar{\zeta})$, $h(u,r)$, $\sigma(u,r)$ and $\alpha(u,r)$ to be determined by the coupled equations  above.

\subsection{Vacuum Einstein Solutions}

\noindent For the sake of completeness and as a check on our field equations we first
set  $\alpha=1$ and choose $\sigma=0$ to retrieve the well known solutions of the vacuum Einstein field equations. Then we must take $C(u,r)$ only\footnote{ This case corresponds to subclass of only Petrov type D solutions as also confirmed by equation (33) which means that the 2-surfaces spanned by complex coordinates have constant curvature, so if they should be compact they are necessarily spheres. The choice of meromorphic functions in (34)  does not provide any new solutions. The same applies to the line element (35) as well.} and $h(u,r)=r e^{-\eta(u)}$  
for some arbitrary function $\eta(u)$. 
Then  the above field equations reduce to the 
following coupled system of equations:
\ba
\frac{1}{r^2}C + \frac{1}{r}C_{r} - \frac{1}{r^2} (  P P_{\zeta \bar{\zeta}} -  P_{\zeta}P_{\bar{\zeta}}) -\frac{4}{r}\eta^{\prime} = 0, \nonumber
\ea
\ba
4\eta^{\prime \prime} + \frac{4}{r}\eta^{\prime} C -2\eta^{\prime}C_{r} - \frac{2}{r} C_{u} =0, \nonumber
 \ea 
\ba
C_{rr} + \frac{2}{r} C_{r} -\frac{4}{r}\eta^{\prime} =0.
\ea
These equations are solved by 
\ba
C(r,u) = 2\eta^{\prime}(u) r + \gamma_{0} e^{2\eta(u)} +\frac{\gamma_{-1}}{r} e^{3\eta(u)}
\ea
provided the function $P(\zeta,\bar{\zeta})$ satisfies
\ba
 P P_{\zeta \bar{\zeta}} -  P_{\zeta}P_{\bar{\zeta}} = \gamma_0.
\ea
If we let $P=e^B$, it reduces to the Liouville equation $B_{\zeta \bar{\zeta}} = \gamma_0 e^{-2B}.$
Hence a  general solution can be written as
\ba
P(\zeta,\bar{\zeta})=  \frac{1 +  \gamma_{0} |\psi(\zeta)|^2}{ |\psi_{\zeta}|},
\ea
in terms of an arbitrary  meromorphic function $\psi(\zeta)$. In particular, for  the choice  $\eta(u)=0$, the Schwarzschild metric is obtained with  
$\psi(\zeta)=\zeta$ and $\gamma_0 =1, \gamma_{-1} = -2M$ where $M$ is the Schwarzschild mass.
\bigskip

\noindent A family of non-static solutions for $\eta(u) \neq 0$ is given by the metric
\ba
g = -\left (2\eta^{\prime}(u) r + e^{2\eta(u)} -\frac{2M}{r}e^{3\eta(u)} \right ) du^2 + 2 du dr + \frac{4 r^2 e^{-2\eta(u)}}{( 1 + |\zeta|^2)^2} d\zeta d\bar{\zeta}. 
\ea

\subsection{Brans-Dicke  Solutions}

\medskip

\noindent In a milestone paper Janis, Newman and Winicour \cite{janis-newman-winicour,janis-robinson-winicour} have constructed an explicit  metric for the static spherically symmetric solution of the coupled Einstein-massless scalar field equations. They have taken the truncated Schwarzschild metric i.e. the exterior Schwarzschild solution as the physical solution 
with a non-zero scalar charge corresponding to   a point mass. We found the corresponding solution of the Brans-Dicke field equations in the Robinson-Trautman form as given by the metric
\ba
g= e^{2\sigma(r)} \left \{ - \left ( \frac{r -M(\mu+1)}{r +M(\mu-1)} \right )^{\frac{1}{\mu}} du^2 + 2 du dr 
+   \frac{ 4 h^2(r) }{(1+|\zeta|^2)^2}  d\zeta d\bar{\zeta} \right \} , 
\ea 
and the scalar field
\ba
\alpha =  \alpha_0 \left (\frac{r-M(\mu+1)}{r+M(\mu-1)}  \right )^{\frac{A}{2\mu}}
\ea
where
$$
\sigma = {\frac{A}{2\mu}} \ln \left |\frac{r + M(\mu-1)}{r - M(\mu+1)}  \right |$$
and
$$
h^2(r) = ( r+M(\mu-1) )^{1+\frac{1}{\mu}} (r-M(\mu+1) )^{1-\frac{1}{\mu}}
$$
provided 
\ba
(\mu -1 ) (\mu +1) =  \frac{c}{2} A^2 .
\ea
Here $r$ stands for  the standard Schwarzschild coordinate. The Schwarzschild mass $M$ and the scalar charge $A$ are identified in the weak field limit\footnote{The scalar charge $A$ is identified as the coefficient of the leading
 term in the expansion of the scalar field $\alpha$ in powers of $\frac{1}{r}$.In case the scalar charge $A=0$ vanishes, the metric reduces to the  standard Schwarzschild
 metric from which the mass $M$ is identified.}. 

\bigskip

\noindent We further look for a family of non-static solutions for which
\ba
\alpha = e^{-\sigma}.
\ea
This assumption simplifies the field equations to a large extent. We take $C(u,r)$ only and set $P P_{\zeta \bar{\zeta}} -P_{\zeta} P_{\bar{\zeta}}=\gamma_0$ as before.
Then we are left with the following system of equations to solve:
\ba
( h h_r C)_r + (2h h_r)_u =\gamma_0, 
\ea
\ba
2\frac{h_{rr}}{h} +c\left (\frac{\alpha_r}{\alpha}  \right )^2 = 0,
\ea
\ba
\frac{h_r}{h}C_u - \frac{h_u}{h}C_r +2\frac{h_{ru}}{h}C +2 \frac{h_{uu}}{h} +c \left ( \frac{\alpha_u}{\alpha} \right )^2 +c\frac{\alpha_u \alpha_r}{\alpha^2} =0,
\ea
\ba
\frac{1}{h^2} \left ( h^2 C_r \right )_r +4 \frac{h_{ur}}{h} +2c\frac{\alpha_u \alpha_r}{\alpha^2} =0,   
\ea
\ba
c \left \{ (h^2C \frac{\alpha_r}{\alpha})_r + 2 h^2 (\frac{\alpha_r}{\alpha})_u + 2h h_r \frac{\alpha_u}{\alpha} + 2h h_u \frac{\alpha_r}{\alpha}   \right \} =0.
\ea
A solution is found as follows:
\ba 
C(u,r) = \gamma_0 e^{-c_1 u} - c_1 r, \qquad  h(u,r) = \sqrt{ e^{c_1 u} r^2 - e^{- c_1 u} c_0^2}, \nonumber \\ \alpha(u,r) =\alpha_0 \left ( \frac{r +c_0e^{-c_1 u}}{r-c_0e^{-c_1 u} }\right )^{\frac{1}{\sqrt{2c}}},  
\ea
where  $\alpha_0, c_0, c_1$ are arbitrary integration constants.This solution can be related with a special case of the 
Robinson-Trautman solutions of the Einstein-massless scalar field equations found recently by Tahamtan and Svitek\cite{tahamtan-svitek1,tahamtan-svitek2}. In particular in their set-up, one should consider $k(x,y)=0$.


\bigskip

\section{Concluding Remarks}

\noindent We considered here a non-Riemannian description of the Brans-Dicke field equations. 
The fact that the field equations are expressed  in a complex null basis facilitates the discussion of Robinson-Trautman space-times.  
All the relevant formulas are written down explicitly in an appendix for future reference.

We also provided a brief discussion on how a change of picture can be affected by local re-scalings of the metric and the scalar field.
As a check on consistency, we presented the special case of vacuum Einstein solutions for $\alpha=1$ and $\sigma=0$.
Similarly we presented a special family of static solutions in the Brans-Dicke-Jordan  picture that corresponds to the well-known Janis-Newman-Winicour family 
which is customarily given in the Einstein picture.   

Denoting the energy-momentum 3-forms of the scalar field and the space-time torsion 2-forms, we have presented the Brans-Dicke field equations obtained by a first order variation with respect to the co-frames and the scalar field.  We generalized the ansatz by inserting a function $h(u,r)$ and an arbitrary conformal factor $\sigma(u,r)$. Solving the complex connection 1-forms from the Cartan structure equations, we calculated the complex null curvature 2-forms, Einstein 3-forms and Weyl curvature 2-forms.
We checked the known source-
free static and non-static solutions verifying the classical results of Robinson-Trautman. On the other hand, we presented the Brans-Dicke static solution of the massless scalar field derived by Janis-Newman-Winicour and a new class of non-static solutions. We noted at the end that more solutions can be obtained by choosing the relationship between the conformal factor $\sigma(u,r)$ and the scalar field $\alpha(u,r)$.

A propagating class of solutions is also found under the assumption $e^{-\sigma}=\alpha$.  These correspond to a special case of Robinson-Trautman solutions
of the Einstein-massless scalar field equations recently found by
Tahamtan and Svitek\cite{tahamtan-svitek1,tahamtan-svitek2}. They have explicitly  derived a Robinson-Trautman solution coupled to a scalar field and have shown that the solution has a singularity created by the divergence of the scalar field therein.
The explicit solution given is asymptotically flat, contains a black hole and carries non-vanishing scalar charge.
Compared to our metric, we can immediately see that the choice for the matter of convenience that they did lies in the solution of the field equations (26) and (27) with the scaling function taken as $\sigma(u,r)=-ln(\alpha(u,r))$.
The only solution to (26) and (27) with the latter assumption is expressed by two functions $b_{1}$ and $b_{2}$ as $C(u,r,\zeta,\bar{\zeta})=b_{1}(u,\zeta,\bar{\zeta})+b_{2}(u,r)$. 


For future work, we plan to find and investigate further new Robinson-Trautman type solutions of the Brans-Dicke gravity.
There are some such  solutions already in the literature\cite{buchdahl,roberts-1,ayse-durmus,vanzo-zerbini,faraoni}. Some solutions describing spherical gravitational waves in Brans-Dicke gravity are also known\cite{wagoner,lee}. 
However, a fresh physical outlook in both respects can be fruitful.  
In particular, Vaidya type of solutions\cite{vaidya,podolsky-svitek1} in the Brans-Dicke-Jordan picture would be physically interesting in view 
of recent work on algebraically special solutions in AdS/CFT\cite{bakas,reall}.

In Brans-Dicke theory, as we have shown above, the gravitational field equations are modified by the scalar field, however, the motion of a (scalar) test mass 
under the influence of gravitation is still assumed to follow the space-time geodesics specified by the Levi-Civita connection of the metric tensor only.  
It was Dirac who first pointed out that it would be more natural to generate the motion of a (scalar) test mass from a locally scale invariant action. 
In such a case, the  resulting equations of motion of a test mass differ in general from a Brans-Dicke geodesic. In the present framework,  
Dirac's proposal amounts to the fact that the test masses under the influence of Brans-Dicke gravity should follow auto-parallels of the connection with torsion. 
It is a well known fact that geodesic curves and auto-parallel curves are distinct in general. In the absence of spinorial matter, no new physics of the fields may arise from re-formulation  of the field equations as given above. On the other hand, the behaviour of scalar test masses in a space-time geometry with torsion would differ 
from their geodesic motion and hence observations must be analysed to decide between these two alternative approaches. 

An analysis of the equations of motion of a test mass in Robinson-Trautman spacetimes we found would be important.
The auto-parallel trajectories in the non-Riemannian framework in the Brans-Dicke-Jordan picture 
differ in general from the geodesic trajectories postulated by Brans and Dicke themselves. Such a difference may give rise to  profound effects 
on the black hole interpretation of the Robinson-Trautman type solutions\cite{bekenstein1,bekenstein2,cebeci-dereli}.

\section{Acknowledgement}

\noindent  Y.\c{S}.  is grateful to Ko\c{c} University for its hospitality and a partial support. M.A. is supported by the Scientific Research Coordination Unit of Pamukkale University, Project No.2018HZDP036. 

 \newpage
 
 \section*{Appendix A:  Spacetime Geometry with Torsion in a Complex Null Basis}

Since we will be using the complex null co-frame in our calculations extensively, in the following we will summarize the basics.
First of all, we are fixing the relations between the orthonormal and complex null co-frames by
\ba
l = \frac{e^3 + e^0}{\sqrt{2}} \; , \quad n = \frac{e^3 - e^0}{\sqrt{2}} \; , \quad
m = \frac{e^1 + ie^2}{\sqrt{2}} \; , \quad \bar{m} = \frac{e^1 -i e^2}{\sqrt{2}}
\ea
and their inverse by 
\ba
e^0 = \frac{l - n}{\sqrt{2}} \; , \quad e^3 = \frac{l + n}{\sqrt{2}} \; , \quad
e^1 = \frac{\bar{m} + m}{\sqrt{2}} \; , \quad e^2 = -i \frac{m-\bar{m}}{\sqrt{2}}.
\ea
The metric becomes
\ba
g=2l\otimes n + 2m \otimes \bar{m} = -e^0 \otimes e^0 + e^1 \otimes
e^1 + e^2 \otimes e^2 +  e^3 \otimes e^3
\ea
where the bar over a symbol denotes its the complex conjugate.

\vspace{2mm}
\noindent
In terms of the orthonormal quantities, the relation $\iota_a e^b = \delta^b_a$ is written in the complex null basis with the 1-forms as
\ba
\iota_l n = \iota_n l = \iota_m \bar{m} = \iota_m \bar{m} = 1 \quad , \quad \iota_l l = \iota_n n = \iota_m m = \iota_{\bar{m}} \bar{m} = 0.
\ea
The relations between the orthonormal and the null interior product is such that
\ba
\iota_l = \frac{\iota_3 - \iota_0}{\sqrt{2}} \; , \quad \iota_n = \frac{\iota_3 + \iota_0}{\sqrt{2}} \; , \quad
\iota_m = \frac{\iota_1 + i \iota_2}{\sqrt{2}} \; , \quad \iota_{\bar{m}} = \frac{\iota_1 -i \iota_2}{\sqrt{2}}
\ea
and their inverse as follows
\ba
\iota_0 = \frac{-\iota_l + \iota_n}{\sqrt{2}} \; , \quad \iota_3 = \frac{\iota_l + \iota_n}{\sqrt{2}} \; , \quad
\iota_1 = \frac{\iota_{\bar{m}} + \iota_m}{\sqrt{2}} \; , \quad \iota_2 = -i\frac{\iota_{m} - \iota_{\bar{m}}}{\sqrt{2}}.
\ea

\vspace{2mm}
\noindent
The Hodge dual defined in the orthonormal quantities is
\ba
*e^{a_1 \cdots a_p} = \frac{1}{(D-p)!} \epsilon^{a_1 \cdots
	a_p}{}_{a_{p+1} \cdots a_D} e^{a_{p+1} \cdots a_D},
\ea
where we consider a $D$ dimensional manifold and $\epsilon_{01 \cdots	D}=+1$ is the totally antisymmetric Levi-Civita symbol in D-dimensions.
In $D=4$ dimensions,  this linear map acting on the null basis 1-forms gives 
\ba
 *1 = i l \w n \w m \w \bar{m}, \nonumber \\  
{}*n = i n \w m \w \bar{m} \; , \;\; {}*l = -i l \w m \w \bar{m} \; , \;\; {} *m = -i l \w n \w m \; , \nonumber \\ 
{}*(l \w n) = -i m \w \bar{m} \; , 
*(l \w m) = -i l\w m \; , 
*(n \w m) = i n \w m .
\ea

\bigskip

\noindent Let us define the  complex null connection 1-forms as
\ba
\omega^k = -\frac{1}{2} (i\omega^0{}_k + \frac{1}{2} \epsilon_{ijk} \omega^{i}_{j})
\ea
where $i,j,k, \cdots =1,2,3$ and $\epsilon_{ijk}$ is totally antisymmetric and satisfies the relation $\epsilon_{123}=+1$. Let's write explicitly
\ba
\omega_1 = -\frac{1}{2} (i\omega^0{}_1 + \omega^2{}_3) \; , \;
\omega_2 = -\frac{1}{2} (i\omega^0{}_2 - \omega^1{}_3) \; , \;
\omega_3 = -\frac{1}{2} (i\omega^0{}_3 + i \omega^1{}_2)
\ea
and conversely
\ba
\begin{array}{lll}
	\omega^0{}_1 = i(\omega_1 -  \bar{\omega}_1 ) \; , & \omega^0{}_2 = i(\omega_2 - \bar{\omega}_2 )\; , & \omega^0{}_3 = i(\omega_3 - \bar{\omega}_3 )\; , \\
	\omega^2{}_3 = -(\omega_1 + \bar{\omega}_1 ) \; ,& \omega^1{}_3 = (\omega_2 + \bar{\omega}_2 ) \; , &  \omega^1{}_2 = -(\omega_3 + \bar{\omega}_3 ). \\
\end{array}
\ea
At this point we adopt a more useful notation by setting
\ba
\omega_+ = \omega^1 + i\omega^2 \; , \;\;
\omega_- = \omega^1 - i\omega^2 \; , \;\;
\omega_0 = \omega^3
\ea
and conversely
\ba
\omega^1 = \frac{1}{2} (\omega_+ + \omega_-) \; , \;\;
\omega^2 = -\frac{i}{2} (\omega_+ - \omega_-) \; , \;\;
\omega^3 = \omega_0.
\ea
Given the torsion 2-forms $T^a = de^a + \omega^a{}_b \wedge e^b$, the first set of Cartan structure equations are written  in terms of the complex quantities as
\ba
 \d l +i(\omega_0 - \bar{\omega}_0) \w l +i \omega_+ \w \bar{m} - i\bar{\omega}_+ \w m  &=& \frac{T^3+T^0}{\sqrt{2}} \;, \nonumber  \\
 \d n -i(\omega_0 - \bar{\omega}_0) \w n + i\bar{\omega}_- \w \bar{m} - i\omega_- \w m  &=& \frac{T^3-T^0}{\sqrt{2}} \;, \nonumber \\
\d m +i(\omega_0 + \bar{\omega}_0) \w m - i\omega_+ \w n - i\bar{\omega}_- \w l  &=&  \frac{T^1+iT^2}{\sqrt{2}} . 
\ea
The corresponding complex null curvature 2-forms are given by
\ba
R^k = -\frac{1}{2} (iR^0{}_k + \frac{1}{2} \epsilon_{ijk} R^{i}_{j}).
\ea
Again a more convenient notation can be introduced so that
\ba
R_+ = R^1 + iR^2 \; , \;\;
R_- = R^1 - iR^2 \; , \;\;
R_0 = R^3
\ea
and conversely
\ba
R^1 = \frac{1}{2} (R_+ + R_-) \; , \;\;
R^2 = -\frac{i}{2} (R_+ - R_-) \; , \;\;
R^3 = R_0.
\ea
Thus the complex null curvature 2-forms are determined by the second set of Cartan structure equations 
\ba
& & R_0 = \d \omega_0 - i\omega_- \w \omega_+ \; , \nonumber \\
& &  R_+ = \d \omega_+ + 2i \omega_0 \w \omega_+ \; , \nonumber \\
& & R_- = \d \omega_- - 2i \omega_0 \w \omega_- \label{eq:egrilik}.
\ea
The Einstein 3-forms $G_a := -\frac{1}{2} R^{bc} \w *e_{abc}$ can be stated now as such:
\ba
\frac{G_3+G_0}{\sqrt{2}} = (R_0 +\bar{R}_0) \w n + R_+ \w m +  \bar{R}_- \w \bar{m}\;, \nonumber \\
\frac{G_3-G_0}{\sqrt{2}} = -(R_0 +\bar{R}_0) \w l -  \bar{R}_+ \w m -  R_+ \w \bar{m}\;, \nonumber \\
\frac{G_1+iG_2}{\sqrt{2}} = -(R_0 -\bar{R}_0) \w m -  \bar{R}_- \w l +  R_+ \w n\;,  \nonumber \\
\frac{G_1-iG_2}{\sqrt{2}} =  (R_0 -\bar{R}_0) \w \bar{m} -  R_- \w l +  \bar{R}_+ \w n\label{eq:Einstein3forms1}.
\ea
\noindent We  also introduce the Weyl curvature 2-forms as usual:
\ba
C_{ab}=R_{ab}-\frac{1}{2}(e_a \w \mathcal{P}_b - e_b \w \mathcal{P}_a) + \frac{{\mathcal{R}}}{6}e_{ab}
\ea
where $R_{ab}$, $\mathcal{P}_a = \iota_{b}R^{b}_{\;\;a} $ and $\mathcal{R}$ are, respectively, the Riemann curvature 2-forms, Ricci 1-forms and the curvature scalar.
In terms of the Schouten 1-forms given by
$$
2 {\mathcal{S}}_a = {\mathcal{P}}_a - \frac{\mathcal{R}}{6} e_a
$$
in $D=4$ dimensions, we have
$$
R_{ab} = C_{ab} + ({\mathcal{S}}_a \w e_b - {\mathcal{S}}_b \w e_a).
$$
Then the  Cotton curvature 2-forms are  determined by
$Y_a = D{\mathcal{S}}_a. $
In the complex null basis, the components of the Weyl curvature 2-forms are given as follows: 
\ba
iC_+&=& iR_{+} + \frac{1}{2}\Bigg(l \w (\frac{{\mathcal{S}}_1 + i {\mathcal{S}}_2}{\sqrt{2}}) -m\w (\frac{{\mathcal{S}}_3 - {\mathcal{S}}_0}{\sqrt{2}})\Bigg) ;\nonumber \\
iC_0&=& iR_{0} +\frac{1}{4}\Bigg(l \w (\frac{{\mathcal{S}}_3 + {\mathcal{S}}_0}{\sqrt{2}}) -n\w (\frac{{\mathcal{S}}_3 - {\mathcal{S}}_0}{\sqrt{2}})\nonumber \\
&&+ m \w (\frac{{\mathcal{S}}_1 - i {\mathcal{S}}_2}{\sqrt{2}})-\bar{m} \w (\frac{{\mathcal{S}}_1 + i {\mathcal{S}}_2}{\sqrt{2}})\Bigg), \nonumber \\
iC_-&=& i R_{-}  + \frac{1}{2}\Bigg(-n \w (\frac{{\mathcal{S}}_1 - i {\mathcal{S}}_2}{\sqrt{2}}) +\bar{m} \w (\frac{{\mathcal{S}}_3 + {\mathcal{S}}_0}{\sqrt{2}})\Bigg).
\ea

\bigskip

 \section*{Appendix B:  Connections and Curvatures in Robinson-Trautman Spacetimes }

From this point on the  Appendix will be used to introduce the complex null co-frame formalism and to illustrate our results.
For the Robinson-Trautman metric (\ref{eq:R-Tmetrik0}), the complex null basis 1-forms in terms of the coordinate 1-forms are given by 
 \ba
l= e^{\sigma}{du} \; , \quad n = e^{\sigma} (dr -
\frac{C}{2} du) \; , \quad m= e^{\sigma} \frac{\sqrt{2}h}{P} d\zeta.
\ea
The torsion 2-forms in complex null basis are found to be 
\ba
\frac{T^3 +T^0}{\sqrt{2}} = e^{-\sigma}\frac{\alpha_r}{\alpha} l \w n, \qquad \frac{T^3 -T^0}{\sqrt{2}} = e^{-\sigma}( \frac{\alpha_u}{\alpha} +\frac{C\alpha_r}{2\alpha}) n \w l, \nonumber \\
\frac{T^1 +iT^2}{\sqrt{2}} = e^{-\sigma}( \frac{\alpha_u}{\alpha} +\frac{C\alpha_r}{2\alpha}) m \w l + e^{-\sigma}\frac{\alpha_r}{\alpha} m \w n.
\ea


\vspace{2mm}
\noindent
After some lengthy calculations we determine the complex null  connection 1-forms with torsion as
\ba
i\omega_+ &=& - e^{-\sigma}\Bigg(\frac{h_{r}}{h}+\sigma_r+\frac{\alpha_{r}}{\alpha}\Bigg)m \; , \\
 i\omega_0 &=& \frac{e^{-\sigma}}{2}\left(\frac{C_r}{2} + \sigma_{u} +\frac{\sigma_{r}C}{2} + \frac{\alpha_{u}}{\alpha} +\frac{\alpha_{r}C}{2\alpha}\right) l -\frac{e^{-\sigma}}{2}\Bigg(\frac{\alpha_r}{\alpha} + \sigma_r\Bigg)n \nonumber \\ & &- e^{-\sigma} \frac{P_{\zeta}}{2\sqrt{2}h} m + e^{-\sigma} \frac{P_{\bar{\zeta}}}{2\sqrt{2}h}\bar{m}, \nonumber \\
i\omega_- &=&  e^{-\sigma} \left( \frac{h_{u}}{h} +\frac{h_{r}C}{2h} + \sigma_{u} +\frac{\sigma_{r}C}{2} + \frac{\alpha_{u}}{\alpha} +\frac{\alpha_{r}C}{2\alpha} \right) \bar{m} + e^{-\sigma} \frac{C_{\zeta}P}{2\sqrt{2}h} l\; .  \nonumber
\ea

\vspace{2mm}
\noindent
Using these in the Cartan structure equations (\ref{eq:egrilik}), we calculate the corresponding complex null curvature 2-forms
\ba
iR_+ &=&-e^{-2\sigma}\Bigg(\frac{1}{2}\frac{\alpha_{rr}C}{\alpha}+\frac{\sigma_r h_u}{h}+\frac{\alpha_r h_r C}{h\alpha}+\sigma_{ur}+\frac{\sigma_r \alpha_u}{\alpha}+\frac{\alpha_r \sigma_u}{\alpha}+\frac{1}{2}\frac{C_r \alpha_r}{\alpha} \nonumber \\
&&+\frac{1}{2}\frac{h_{rr}C}{h}+\frac{\sigma_rC\alpha_r}{\alpha}+\frac{\alpha_r h_u}{h\alpha}+\frac{\alpha_u h_r}{h\alpha}+\frac{\sigma_r h_r C}{h}+\frac{\alpha_{ur}}{\alpha}+\frac{h_{ur}}{h} \nonumber \\
&&+\frac{1}{2}\sigma_{rr}C+\frac{\sigma_u h_r}{h}+\frac{1}{2}\frac{h_r C_r}{h}+\sigma_r \sigma_u+\frac{1}{2}\sigma^2_r C +\frac{1}{2}\sigma_r C_r\Bigg)\hspace{2mm}l \w m\nonumber \\
&&+e^{-2\sigma}\Bigg(\frac{2\sigma_r \alpha_r}{\alpha}-\frac{\alpha_{rr}}{\alpha}-\sigma_{rr}+\sigma^2_r-\frac{h_{rr}}{h}+\frac{2\alpha^2_r}{\alpha^2}\Bigg) \hspace{2mm}n \w m\; , \nonumber \\
iR_0 &=&e^{-2\sigma}\Bigg(\frac{1}{2}\frac{\alpha^2_rC}{\alpha^2}-\frac{1}{4}C_{rr}-\frac{1}{2}\sigma_rC_r+\frac{\alpha_u\alpha_r}{\alpha^2}-\frac{\alpha_{ur}}{\alpha}-\frac{1}{2}\sigma_{rr}C-\frac{1}{2}\frac{C_r\alpha_r}{\alpha}\nonumber \\
&&-\frac{1}{2}\frac{\alpha_{rr}C}{\alpha}-\sigma_{ur}\Bigg)\hspace{2mm}l \w n\hspace{2mm} + e^{-2\sigma}\Bigg(-\frac{1}{4\sqrt{2}}\frac{C_{r\zeta}P}{h}+\frac{1}{2\sqrt{2}}\frac{Ph_rC_{\zeta}}{h^2}\Bigg) l \w m \nonumber \\
&&-e^{-2\sigma}\Bigg(\frac{1}{2}\sigma^2_rC+\frac{\sigma_r h_rC}{h}+\frac{\sigma_r C \alpha_r}{\alpha}+\frac{\sigma_r h_u}{h}-\frac{\alpha_r h_u}{h\alpha}+\frac{1}{2}\frac{h^2_rC}{h^2}\nonumber \\
&&-\frac{1}{2}\frac{P_{\zeta\bar{\zeta}}P}{h^2}+\frac{1}{2}\frac{P_{\zeta }P_{\bar{\zeta}}}{h^2}+\frac{h_r h_u}{h^2}+\frac{1}{2}\frac{\alpha^2_rC}{\alpha^2}+\frac{\sigma_r \alpha_u}{\alpha}-\frac{\alpha_u h_r}{h\alpha}+\frac{\alpha_r\alpha_u}{\alpha^2}\nonumber \\
&&+\frac{\sigma_u h_r}{h}+\sigma_r \sigma_u+\frac{\alpha_r \sigma_u}{\alpha}+\frac{\alpha_r h_r C}{h\alpha}\Bigg)\hspace{2mm} m \w \bar{m}\nonumber\\
&&-e^{-2\sigma}\Bigg(\frac{1}{4\sqrt{2}}\frac{PC_{r\bar{\zeta}}}{h}+\frac{1}{2\sqrt{2}}\frac{P\sigma_rC_{\bar{\zeta}}}{h}+\frac{1}{2\sqrt{2}}\frac{\alpha_rC_{\bar{\zeta}}P}{h\alpha}\Bigg)\hspace{2mm} l \w \bar{m}, \nonumber \\
iR_- &=&e^{-2\sigma}\Bigg(-\frac{h_uC_r}{2h}-\frac{\alpha_u C_r}{2\alpha}-\frac{2\sigma_u \alpha_u}{\alpha}-\sigma_u\sigma_rC-\frac{2\alpha_u C \alpha_r}{\alpha^2}-\frac{\sigma_uC\alpha_r}{\alpha}\nonumber \\
&&-\frac{\sigma_rC\alpha_u}{\alpha}-\frac{\sigma_rC^2\alpha_r}{2\alpha}+\sigma_{ur}C+\frac{C\alpha_{ur}}{\alpha}-\frac{2\alpha^2_u}{\alpha^2}+\sigma_{uu}-\sigma^2_u\nonumber \\
&&+\frac{\alpha_{uu}}{\alpha}+\frac{h_{uu}}{h}+\frac{h_{ur}C}{h}+\frac{h_rC_u}{2h}-\frac{1}{2}\sigma_uC_r+\frac{1}{4}\sigma_{rr}C^2+\frac{1}{4}\frac{\alpha_{rr}C^2}{\alpha^2}\nonumber \\
&&-\frac{1}{4}\sigma^2_rC^2+\frac{1}{4}\frac{h_{rr}C^2}{h}-\frac{1}{2}\frac{\alpha^2_rC^2}{\alpha^2}+\frac{1}{2}\frac{C_u\alpha_r}{\alpha}+\frac{1}{2}\sigma_rC_u-\frac{1}{4}\frac{P^2C_{\zeta\bar{\zeta}}}{h^2}\Bigg)\hspace{2mm} l \w \bar{m}\nonumber \\
&&+e^{-2\sigma}\Bigg(\frac{h_{ur}}{h}+\frac{\alpha_{ur}}{\alpha}+\sigma_r\sigma_u+\frac{\sigma_r h_r C}{h}+\frac{\sigma_rC\alpha_r}{\alpha}+\frac{\sigma_r\alpha_u}{\alpha}+\frac{1}{2}\sigma_rC_r\nonumber \\
&&+\frac{\alpha_r h_r C}{h\alpha}+\sigma_{ur}+\frac{\alpha_r h_u}{h\alpha}+\frac{\sigma_u h_r}{h}+\frac{1}{2}\sigma_{rr}C+\frac{1}{2}\sigma^2_rC+\frac{1}{2}\frac{h_rC_r}{h}\nonumber \\
&&+\frac{1}{2}\frac{h_{rr}C}{h}+\frac{\sigma_r h_u}{h}+\frac{\alpha_r\sigma_u}{\alpha}+\frac{1}{2}\frac{C_r\alpha_r}{\alpha}+\frac{\alpha_u h_r}{h\alpha}+\frac{1}{2}\frac{\alpha_{rr}C}{\alpha}\Bigg) \hspace{2mm} n\w \bar{m}\nonumber  \\
&&+e^{-2\sigma}\Bigg(\frac{1}{2\sqrt{2}}\frac{Ph_{r}C_{\zeta}}{h^2}-\frac{1}{2\sqrt{2}}\frac{P\sigma_{r}C_{\zeta}}{h}-\frac{1}{2\sqrt{2}}\frac{C_{r\zeta}P}{h}-\frac{1}{2\sqrt{2}}\frac{C_{\zeta}P\alpha_r}{h\alpha}\Bigg)\hspace{2mm} l \w n \nonumber\\
&&-e^{-2\sigma}\Bigg(\frac{1}{4}\frac{P^2C_{\zeta\zeta}}{h^2}+\frac{1}{2}\frac{PC_{\zeta}P_{\zeta}}{h^2}\Bigg)\hspace{2mm} l \w m \nonumber\\
&&+e^{-2\sigma}\Bigg(\frac{1}{2\sqrt{2}}\frac{P\sigma_{r}C_{\zeta}}{h}+\frac{1}{2\sqrt{2}}\frac{PC_{\zeta}\alpha_{r}}{h\alpha}+\frac{1}{2\sqrt{2}}\frac{Ph_{r}C_{\zeta}}{h^2}\Bigg)\hspace{2mm} m \w \bar{m}\; . 
\ea

\noindent Finally, the Weyl curvature 2-forms are
\ba
C_+&=& -\frac{1}{6}\Bigg(-\frac{P_{\zeta\bar{\zeta}}P}{h^2}+\frac{P_{\zeta}P_{\bar{\zeta}}}{h^2}+\frac{h_r^2C}{h^2}-\frac{h_rC_r}{h}-\frac{h_{rr}C}{h}-2\frac{h_{ur}}{h}+2\frac{h_uh_r}{h^2} \nonumber\\
&&+\frac{C_{rr}}{2}\Bigg)e^{-2\sigma} \hspace{1mm} il \w m,\nonumber \\
C_-&=& \frac{P}{2h^2}\Bigg(P_{\zeta}C_{\zeta}+\frac{PC_{\zeta\zeta}}{2}\Bigg)e^{-2\sigma} \hspace{1mm} il \w m \nonumber \\
&& +\frac{1}{6}\Bigg(-\frac{P_{\zeta\bar{\zeta}}P}{h^2}+\frac{P_{\zeta}P_{\bar{\zeta}}}{h^2}+\frac{h_r^2C}{h^2}-\frac{h_rC_r}{h}-\frac{h_{rr}C}{h}-2\frac{h_{ur}}{h}+2\frac{h_uh_r}{h^2} \nonumber\\
&&+\frac{C_{rr}}{2}\Bigg)e^{-2\sigma} \hspace{1mm} in \w \bar{m} -\frac{P}{2h}\Bigg(\frac{h_rC_{\zeta}}{h}-\frac{C_{r\zeta}}{2}\Bigg)e^{-2\sigma} \hspace{1mm} \frac{i}{\sqrt{2}}(l \w n + m \w \bar{m}),\nonumber \\
C_0&=&-\frac{P}{2\sqrt{2}h}\Bigg(\frac{h_rC_{\zeta}}{h}-\frac{C_{r\zeta}}{2}\Bigg)e^{-2\sigma} \hspace{1mm} il \w m \nonumber\\
&&+\frac{1}{3\sqrt{2}}\Bigg(-\frac{P_{\zeta\bar{\zeta}}P}{h^2}+\frac{P_{\zeta}P_{\bar{\zeta}}}{h^2}+\frac{h_r^2C}{h^2}-\frac{h_rC_r}{h}-\frac{h_{rr}C}{h}-2\frac{h_{ur}}{h}+2\frac{h_uh_r}{h^2} \nonumber\\
&&+\frac{C_{rr}}{2}\Bigg)e^{-2\sigma} \hspace{1mm} \frac{i}{\sqrt{2}}(l \w n + m \w \bar{m}).
\ea
It turns out from the above expressions that the Weyl scalars $\Psi_0 = \Psi_1=0$ vanish so that our spacetime is of Petrov type II in general.

\medskip

\noindent In order to determine the coupled   equations to be solved we also write down the stress-energy-momentum tensor of the scalar field in the complex null basis as
follows: 
\ba
\frac{\tau_3+\tau_0}{\sqrt{2}}=-e^{-2\sigma}(\alpha_u+\frac{\alpha_rC}{2})^2\hspace{1mm} il \w m \w \bar{m}, \quad 
\frac{\tau_3-\tau_0}{\sqrt{2}}
 =e^{-2\sigma}\alpha^2_r \hspace{1mm} in \w m \w \bar{m}, \nonumber
\ea
\ba
\frac{\tau_1+i\tau_2}{\sqrt{2}} = e^{-2\sigma}\alpha_r(\alpha_u+\frac{\alpha_rC}{2})\hspace{1mm} il \w n \w m.
\ea

\newpage

\end{document}